\documentclass[conference]{IEEEtran}
\IEEEoverridecommandlockouts
\usepackage{cite}
\usepackage{amsmath,amssymb,amsfonts}
\usepackage{algorithmic}
\usepackage{graphicx}
\usepackage{textcomp}
\usepackage{amsmath}
\usepackage[version=4]{mhchem}
\usepackage{siunitx}

\usepackage{multicol}
\usepackage{amssymb}
\usepackage{multirow}
\usepackage{verbatim}
\usepackage{graphicx}

\usepackage{subcaption}
\usepackage{epstopdf}
\usepackage{tabularx}
\usepackage{floatrow}
\usepackage{booktabs} % \**rule table line package
\usepackage{longtable,tabularx}
\usepackage{arydshln} % dashline package
\usepackage{url}
\usepackage{hyperref} % makes references hyperlinks
\usepackage{textcomp} % Symbol Expansion Package

\usepackage{epstopdf}
\usepackage{xcolor}
\def\BibTeX{{\rm B\kern-.05em{\sc i\kern-.025em b}\kern-.08em
    T\kern-.1667em\lower.7ex\hbox{E}\kern-.125emX}}
\begin{document}

\title{Graph Learning Based Decision Support for Multi-Aircraft Take-Off and Landing at Urban Air Mobility Vertiports\\

\author{Prajit KrisshnaKumar$^{1*}$, Jhoel Witter$^{1*}$, Steve Paul$^{1}$, Karthick Dantu$^{2}$, Souma Chowdhury$^{1}$
\thanks{$^{1}$ Department of Mechanical and Aerospace Engineering, University at Buffalo, Buffalo,
NY, 14260 USA.}%
\thanks{$^{2}$Department of Computer Science and Engineering, University at Buffalo, Buffalo,
NY, 14260 USA.}%
\thanks{$^\dagger$ Corresponding Author, 
{\tt\small soumacho@buffalo.edu}}
\thanks{This work was supported by Stephen Still Institute for Sustainable Transportation and Logistics (SSISTL). Any opinions, findings, conclusions, or recommendations expressed in this paper are those of the authors and do not necessarily reflect the views of the SSISTL. }
}
 \thanks{
 *Equal Contribution by Authors
}
}
\maketitle

\begin{abstract}
Majority of aircraft under the Urban Air Mobility (UAM) concept are expected to be of the electric vertical takeoff and landing (eVTOL) vehicle type, which will operate out of vertiports. While this is akin to the relationship between general aviation aircraft and airports, the conceived location of vertiports within dense urban environments presents unique challenges in managing the air traffic served by a vertiport. This challenge becomes pronounced within increasing frequency of scheduled landings and take-offs. This paper assumes a centralized air traffic controller (ATC) to explore the performance of a new AI driven ATC approach to manage the eVTOLs served by the vertiport. Minimum separation-driven safety and delays are the two important considerations in this case. The ATC problem is modeled as a task allocation problem, and uncertainties due to communication disruptions (e.g., poor link quality) and inclement weather (e.g., high gust effects) are added as a small probability of action failures. To learn the vertiport ATC policy, a novel graph-based reinforcement learning (RL) solution called “Urban Air Mobility- Vertiport Schedule Management (UAM-VSM)" is developed. This approach uses graph convolutional networks (GCNs) to abstract the vertiport space and eVTOL space as graphs, and aggregate information for a centralized ATC agent to help generalize the environment. Unreal Engine combined with Airsim is used as the simulation environment over which training and testing occurs. Uncertainties are considered only during testing, due to the high cost of Mc sampling over such realistic simulations. The proposed graph RL method demonstrates significantly better performance on the test scenarios when compared against a feasible random decision-making baseline and a first come first serve (FCFS) baseline, including the ability to generalize to unseen scenarios and with uncertainties.
\end{abstract}

\section{Introduction}
Technology for transportation is rapidly evolving every day, with self-driving cars and autonomous air package delivery around the corner. It is estimated by 2050 around 68\% of the world's population will live in urban areas \cite{united_nations_2018}. Urban Air Mobility (UAM) adds a new dimension to the mode of transportation, where vertical takeoff and landing (VTOL) devices are used for transporting people at moderate altitudes \cite{rothfeld2020urban}. The concept of UAMs dates back to 1953 when New York Airways operated commercial air taxis using helicopters. With the current advancements in electrical, propulsion, and battery fields, air taxis are becoming more viable and economical \cite{BAURANOV2021100726}. Companies like Uber are racing towards the development and deployment of VTOLs in urban areas \cite{horne2019next}, but the time frame remains a mystery as the deployment faces several challenges including government regulations. Due to the availability of current transportation spaces and the estimated population, UAMs are inevitable in the near future.

Currently, the Air Traffic Control (ATC) operates all the vehicles with the ability to fly \cite{vascik2017constraint}. A vertiport is an area where the VTOLs take-off, land, and charge their batteries \cite{daskilewicz2018progress}. When it comes to VTOLs, the number of vehicles entering or leaving a vertiport will be hundreds to thousands in an hour \cite{guerreiro2020capacity}.  In this case, it is more challenging to control the aircraft's landing/take-off (L/TO) and it raises concerns about safety and regulation. The First-come, First-served concept does not serve well in the presence of uncertainties and emergencies.  In this paper, we propose a solution for regulating the VTOLs inside the vertiport zone while simultaneously maintaining the safety of the VTOLs. This is a multi-dimensional problem and needs to be modeled beyond the simple linear mathematical modeling, hence we created a novel 3D simulation environment incorporated with realistic physics and primitives like path planning. Importantly our simulator runs much faster than real-time and this helps us in collecting the data required for learning. Two different learning-based algorithms are trained and compared with each other. There are several prior works that discuss the modeling of vertiport\cite{preis2022vertiport} and VTOLs \cite{bacchini2019electric}. Here the design of vertiport and VTOLs are out of scope, instead we consider the port to be a helipad and UAV to be the VTOL, since the concept applies the same. 

Recently, Artificial Neural Network (ANN) based learning algorithms are used in various intelligent autonomous system and it plays a better role in decision-making when compared to humans \cite{li2017deep}. Most successful RL applications such as self-driving cars and robotics include more than a single agent and are solved as Multi-Agent Reinforcement learning (MARL) problems\cite{https://doi.org/10.48550/arxiv.1911.10635}. Over the past few years, there have been several notable works on applying Graph-based Reinforcement Learning (RL) for various single and multi-agent Combinatorial Optimization (CO) problems \cite{Kool2019, barrett2019exploratory,doi:10.2514/6.2022-3911, khalil2017learning, Kaempfer2018LearningTM, 9750805, li2018combinatorial, nowak2017note, Tolstaya2020MultiRobotCA, Sykora2020, paul_cvrp, Dai2017, Paul_ICRA}. Here, the state space variables which can be modeled as a graph are encoded using Graph Neural Networks (GNN), which will be part of the policy network. There are several difficulties in MARL such as the curse of dimensionality or the exponential growth in state-action space, Non stationarity- complicated dynamics,  and the credit assignment-the ambiguity on which agent has to be rewarded \cite{hernandez2018multiagent}. Some of the recent works to overcome the limitations involve converting the MARL problems to single agent \cite{behjat2021learning} (centralized training), the experience of all the agents are collected and trained by one agent and with decentralized implementation where the trained model is being implemented on all the agents to enable decentralized decision-making \cite{lyu2021contrasting}. We have formulated our problem as a single agent, though the number of agents can go beyond the numbers in this paper. Furthermore, we discretized a continuous environment and formulated the state space with discrete space which is easier to learn compared to continuous state space. 

The main contributions of this paper are 1) Formulation of the vertiport operations management as a Markov Decision Process (MDP) -- a.k.a. short-term-scale VTOL landing/take-off(L/TO) scheduling problem, 2) Development of 3D simulation environment for modeling the UAM vertiport operations and 3) Development of a (graph) learning framework to provide the policies for timing the L/TO of the aircraft within the vertiports operational space considering environmental and operational uncertainties. The remainder of the paper is organized as follows. In section \ref{sec:problemform} we explain the MDP formulation and the learning approach. In section \ref{modelling} we briefly explain the architecture of the simulation, together with the working of its individual components and the details on the learning algorithm used.  In section \ref{sec:case_studies}, the different case studies are explained. 

\section{Vertiport Operation Management: Formulation and Learning Approach}
\label{sec:problemform}

% The problem being solved is for a single vertiport with different number of ports a.k.a landing pads. Each VTOL will have a scheduled time at which it will visit the main vertiport. Once the VTOL reaches the vertiport it hovers over one of the empty hovering spots in the vertiport zone and communicates with the central server to get information on the available ports. This information is then encoded and transferred to the trained agent onboard, which will then decide the action to be taken. If the VTOL is already landed it has to decide between takeoff and charging. Here the job of the central server is to pass the data and update them based on the actions of the VTOLs. All the decisions are being taken by the individual VTOLs and is decentralized. The actions are high level, i.e., we don't control the motor of VTOLs instead we assign the position or location it has to go each time step and the simulation takes care of path planning. The MDP formulation and learning agents are explained below. 

\begin{figure}
\centering
\includegraphics[width=0.8\linewidth]{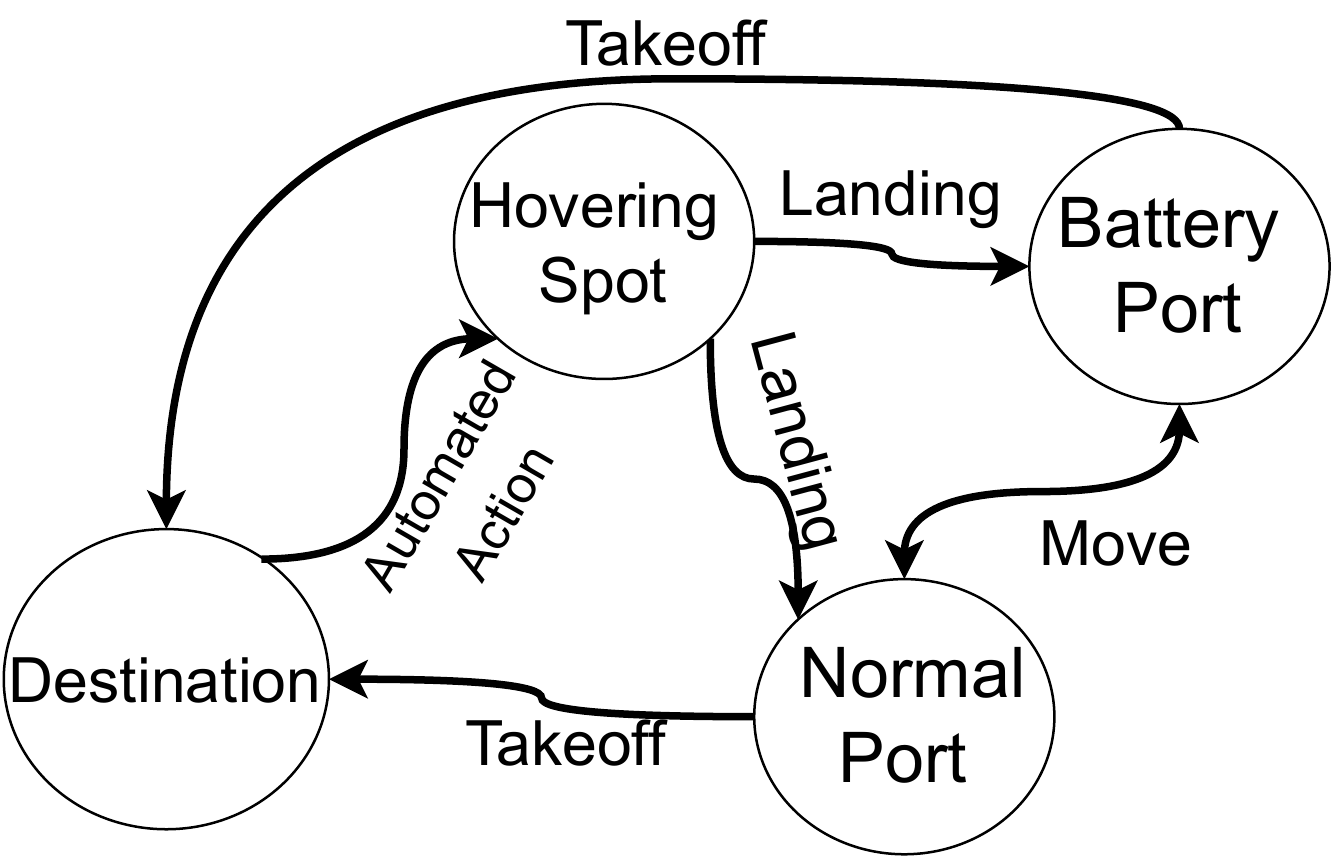}
\caption{State and action diagram for the vertiport environment}
\label{figs:state_action}
\end{figure}

This is an Urban Air Mobility - Vertiport Schedule Management problem, and the goal is to design a GCN policy capable of training an ATC agent. This agent must be able to: allocate tasks to eVTOLs in its airspace (charging, taking off, landing, hovering), maintain high charge levels across all eVTOLs, avoid collisions and follow each eVTOLs specified flight plan. The environment consists of 2 normal ports, 1 charging port, 5 destinations outside of the agent air space, and 7 hovering spots. A simplified movement chart of the environment can be found in figure \ref{figs:state_action}. The following sub-sections will go into more detail about the environment and Markov Decision Process (MDP) formulation we chose for this problem.

\subsection{Environment}
\begin{figure}
\centering
\includegraphics[width=1.15\linewidth]{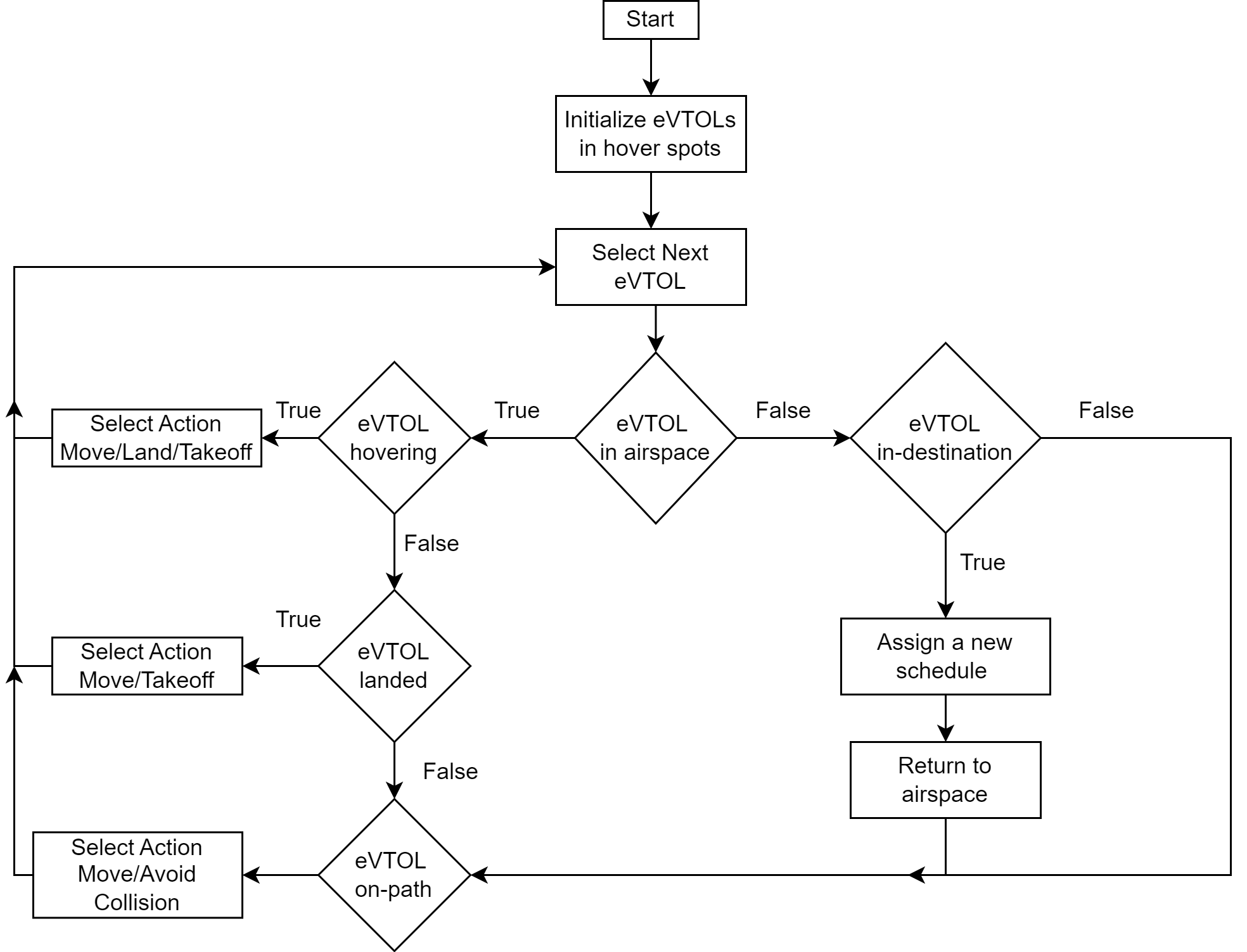}
\caption{Flowchart Representing the Decision-Making} 
\label{figs:algorithm}
\end{figure}

The environment is initialized with 4 eVTOLs, and each one takes off to a random destination, and returns. Each eVTOL receives a new flight plan when they land at the vertiport, which designates a time 10 to 20 minutes in the future where they'll need to take off, and which destination they'll need to go to. Once an eVTOL reaches a destination, they'll receive a flight plan to return to the vertiport automatically, and the ATC will need to land them within 15 minutes of their arrival. This is further explained in figure \ref{figs:algorithm}. Each eVTOL starts off with a full charge and discharges at each step. The discharge rate depends on the state of the eVTOL:

\begin{equation}
\textrm{discharge rate} =
\left\{
	\begin{array}{ll}
		\textrm{distance traveled} \times 0.5  & \textrm{if cruising}\\
            2  & \textrm{if hovering}\\
    	4 &  \textrm{if idling on ground}\\
	\end{array}
\right.
\end{equation}

This is done to make sure the eVTOL will discharge its battery when it's not moving. At every step, an eVTOL can recover 10\% of its battery if it's landed on the charging port. Each E-VTOL moves asynchronously, such that there can be multiple E-VTOLs moving at the same time. Due to this configuration, the agent can learn to avoid collisions with two or more E-VTOLs intersecting. At every time step, the agent will select a new E-VTOL to take an action, and this goes in order. If the selected E-VTOL is currently at a destination, the agent will wait for it to re-enter the vertiport airspace before selecting an action for it to take. The simulation runs at 300x real-time, so every second that passes equates to 5 minutes in the simulator. This is good for training, as it allows for step times of close to 0.2 seconds, or approximately 288 seconds (4.8 minutes) per episode (1440 steps). Each episode is approximately 24 hours in the simulation, a full day of operation. Each E-VTOL is updated with a minimum frequency of $45 Hz$, which includes updating all features (location, delay, battery status, and E-VTOL status).

\subsection{MDP Formulation}
The problem is modeled as a Markov decision process (MDP) which has the following parameters:

\begin{equation}\label{eq:MDP} 
    \begin{aligned}
     &s = \mathcal{F}_s(b_{i}, c_{i}, l_{i}, x_{i, p} , y_{i, p}, p_{a, t} )\\
     &a = \mathcal{F}_a(s)\\
     &r = \mathcal{F}_r(a, s, \S, \beta, \gamma, \tau, \gamma, w_n)\\
    \end{aligned}
\end{equation}
where $i$ and $p$ stand for all the vehicles and ports respectively. The state and action space can be found in table \ref{tab:MDP}, and the reward is shown in equation \ref{eq:reward}.$\tau$ is the takeoff coefficient, $\gamma$ is the landing coefficient, $\lambda$ is the battery coefficient, $\beta$ is the delay coefficient, $\S$ is the safety coefficient, and $w_n$ are the weights. These coefficients are further explained below.

\begin{equation}\label{eq:reward} 
    \begin{aligned}
        R = w_1\tau + w_2\gamma + w_3\lambda + w_4\beta + w_5\S
    \end{aligned}
\end{equation}\\

\subsubsection{Takeoff \& Landing coefficient}

We define a ``\textbf{good}'' takeoff as one where the E-VTOL is: {\bf i)} taking off on time (within 5 minutes of its scheduled takeoff time); {\bf ii)} taking off with a battery level greater than 30\%. This criterion is the same for a ``\textbf{good}'' landing except that the E-VTOL can also choose to land earlier than its scheduled time. Both $\tau$ and $\gamma$ $\in \{-5,5\}$.

\subsubsection{Battery coefficient}
This coefficient is defined as:
\begin{equation}
\lambda = 
\left\{
	\begin{array}{ll}
		5 \times \frac{battery\_remaining}{100} & \mbox{if } battery\_remaining \geq 30\\
        -5  & else\\
	\end{array}
\right.
\end{equation}
where the value is gradually increased as the E-VTOL charges. To further discourage traveling with a critical battery level, there is a penalty assigned once the battery percentage drops below 30\%.
To achieve the battery coefficient the agent would need to keep each E-VTOL fully charged.

\subsubsection{Delay coefficient}
Delay is introduced in the environment once an E-VTOL has missed it's window for either taking off or landing. This delay will rise as the simulation time goes on until the E-VTOL travels and receives a new schedule. The delay coefficient is defined as:
\begin{equation}
\beta = -5+10 \times e^{-delay}
\end{equation}

where the $delay$ term is in minutes. This encourages the agent to keep the delay as low as possible to achieve the maximum delay coefficient.

\subsubsection{Safety coefficient}

\begin{figure}
\centering
\includegraphics[width=0.95\linewidth]{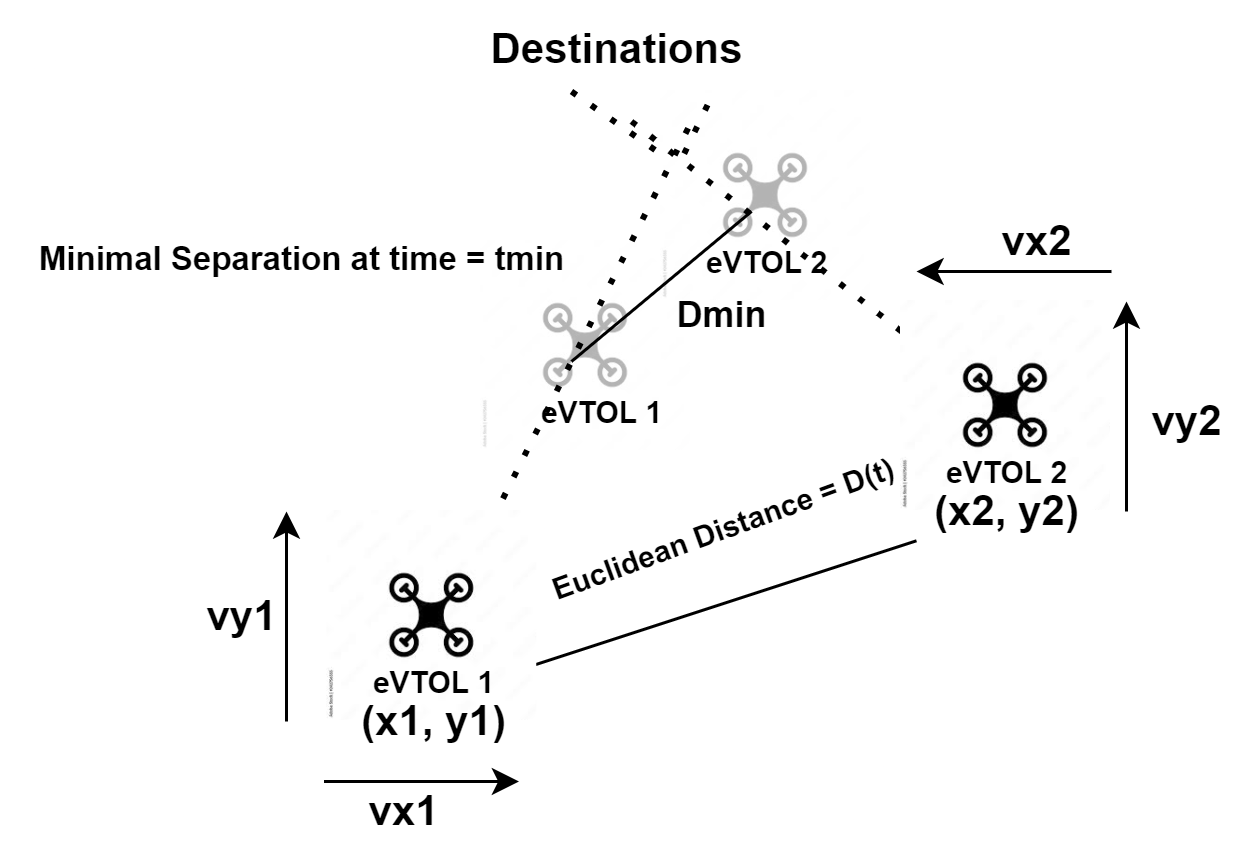}
\caption{Minimal Separation Scenario}
\label{figs:separation}
\end{figure}

Before the safety coefficient can be calculated, the environment will check to see if the selected E-VTOL: {\bf i.)} Is currently en-route to a location; {\bf ii.)} Has an intersecting path with another E-VTOL that is en route. This can be visualized in figure \ref{figs:separation}. In the figure, two E-VTOLs are traveling towards an intersection point. The distance between them at any given point is made into a function of time using the euclidean distance combined with their instantaneous position and velocity vectors:

\begin{equation}\label{eq:euclid} 
D(t) = \sqrt{ (x_1 - x_2 + tv_{x_1} - tv{x_2})^2 + (y_1 - y_2 + tv_{y_1} - tv{y_2})^2 } 
\end{equation}

where $x_n, y_n, v_{x_n}, v_{y_n}$ are the position and velocity components of E-VTOLs 1 and 2. This equation is then differentiated with respect to time and solved for the local minimum, $t_{min}$:

\begin{equation}\label{eq:tmin} 
t_{min} =  -\frac{2(v_{x_1} - v_{x_2})(x_1 - x_2) + 2(v_{y_1} - v_{y_2})(y_1 - y_2)}{2(v_{x_1} - v_{x_2})^2 + 2(v_{y_1} - v_{y_2})^2}
\end{equation}

$t_{min}$ is then plugged back into equation \ref{eq:euclid} to get the minimum separation $D_{min}$. Each simulated E-VTOL has an occupant space of 1x1 meters, so if $D_{min}$ is less than 3 meters and the agent doesn't take evasive action, the agent will be penalized:

\begin{equation}
\S =
\left\{
	\begin{array}{ll}
		0  & \mbox{if } D_{min}\ is\ None\\
		-5 & \mbox{if } D_{min} \leq 3.0\ \&\ action \neq avoid\ collision\\
		5 & \mbox{if } D_{min} \leq 3.0\ \&\ action = avoid\ collision\\
	\end{array}
\right.
\end{equation}

\subsubsection{Reward weights}
Each coefficient is multiplied by a weight $w_1, w_2,..w_n$ based on the importance of each coefficient. In our problem, safety is considered more important, and maximum weight is allotted for safety coefficient $\S$ followed by $\beta, \tau, \gamma, \lambda$.

% For our training, 3 vertiports and 4 VTOLS are considered. At the beginning of each episode, all the vehicles are assigned a random schedule and location. We iterate through the vehicles and select the one which is idle(either in hover spot or in landing port). Our learning algorithm predicts an action for that vehicle based on its status.If the action is landing, the VTOL chooses a port based on the safety parameter and lands there.  If the action is takeoff, The VTOL goes to the destination according to its schedule, Once it reach the destination, we assign a new schedule and the vehicle returns to the main port. All vehicles occupy one of the empty hover spots and wait till the iteration reaches it.  The training episode terminates when one of the following conditions is satisfied, 

% \begin{figure}
% \centering
% \includegraphics[width=0.4\linewidth]{Figs/rewardplot.png}
% \caption{Reward Distribution Plot for Takeoff Scenarios } 
% \label{figs:reward3d}\vspace{-0.2cm}
% \end{figure}

% \begin{itemize}
%   \item $m>1000$
%   \item $\sum_{i=1}^{n} T_{n} > 2000$
%   \item $\zeta < -10 $
% \end{itemize}

% At every time step, the VTOLs are provided with 2 alternative paths to be taken with a Boolean value indicating the safety. The agent has to decide the best path it must take to avoid collision. More details on how these paths are chosen will be explained in final version. The state space variations in the learning algorithms are explained below. 

\begin{table}\label{tab:mdp_formulation}
\vspace{-2mm}
\footnotesize
\centering
\begin{tabular}{|l|l|} 
\hline
Type                              & Variable                        \\ 
\hline
\multirow{3}{*}{Vertiport states} & Availability - $P_a$            \\
                                  & Port type - $P_t$               \\
                                  & Location - $(x_p, y_p)$         \\ 
\hline
\multirow{5}{*}{VTOL states}     & Current status - $c_i$           \\
                                  & Battery capacity - $b_i$        \\
                                  & Schedule status - $l_i$         \\
                                  & Location - $(x_i, y_i)$         \\
                                  & Port availability - $P_a$       \\
\hline
\multirow{5}{*}{Action Space}     & Stay still                       \\
                                  & Takeoff                          \\
                                  & Move/ land in normal port - 1,2  \\
                                  & Move/ land in battery port - 1   \\
                                  & Move to hover spots - 1,2,3,4,5,6,7    \\
                                  & Continue previous action        \\
                                  & Avoid collision                 \\
\hline
\end{tabular}
\caption{MDP formulation}
\label{tab:MDP}
\vspace{-2mm}
\end{table}

\section{Learning Architecture}
\label{sec:learning}
This paper focuses on a deep reinforcement learning framework known as proximal policy optimization (PPO) \cite{schulman2017proximal}. PPO is similar to Trust Region Policy Optimization, otherwise known as TRPO \cite{https://doi.org/10.48550/arxiv.1502.05477}. What sets PPO apart is the ability to clip policy expansion through the use of its unique objective function, which allows for safer policy exploration without the cost of larger unstable policy updates. This clipping parameter can be increased or decreased to control how big the updates are, and when combined with Adams gradient descent it makes for a powerful learning algorithm. We use OpenAI Gym and Stable Baselines 3 \cite{openai} \cite{stable-baselines} for reinforcement learning, and the general flow of the training environment can be found in figure \ref{figs:flow}.

\subsection{GCN Agent}

\begin{figure}
\centering
\includegraphics[width=0.95\linewidth]{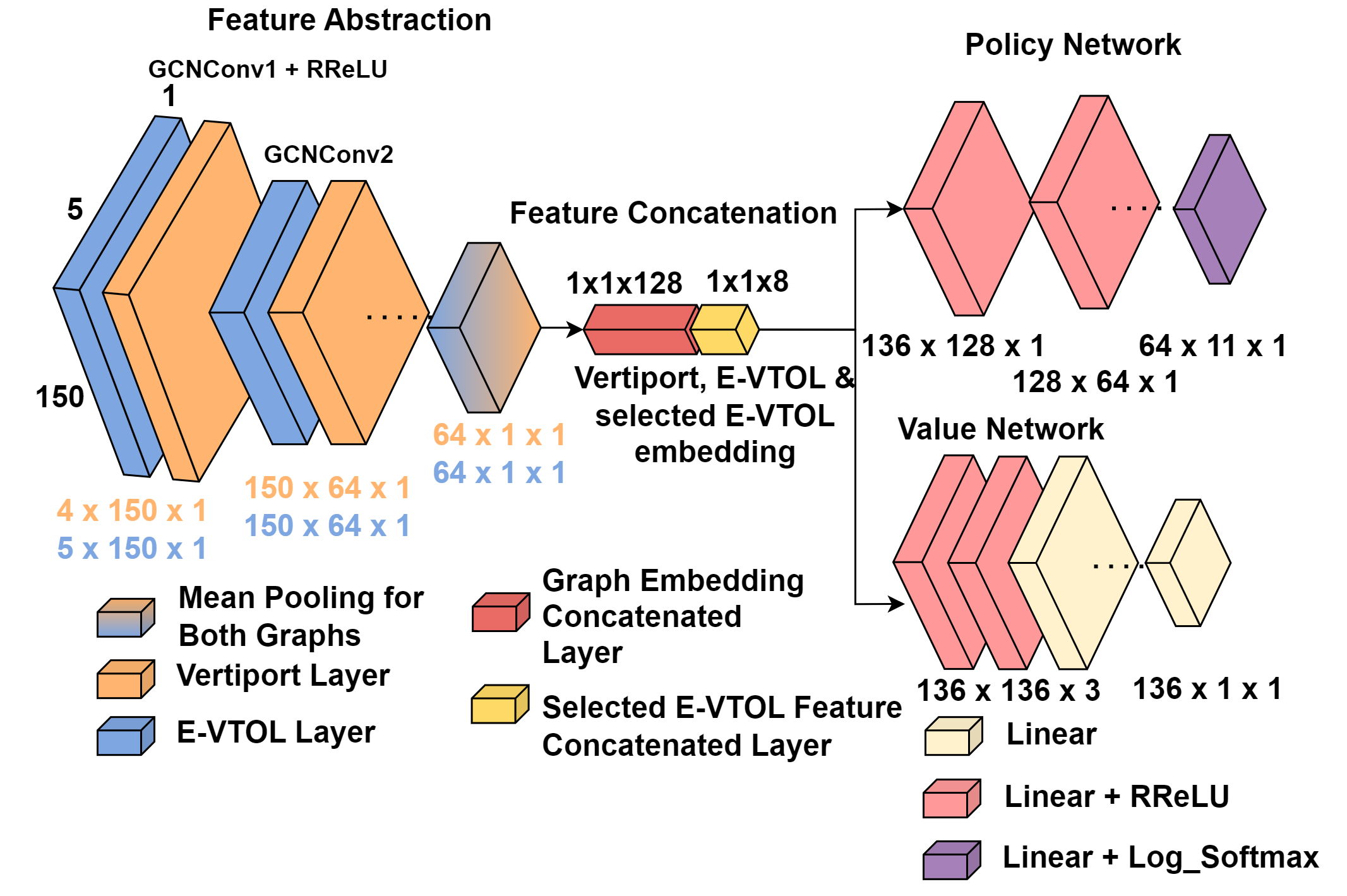}
\caption{Feature abstraction, value and policy networks for the proposed GRL agent}
\label{fig:grlagentnetwork}\vspace{-0.2cm}
\end{figure}

As stated above, PPO is a state-of-the-art actor-critic RL method that has demonstrated high efficiency, wide adaptability, and robust reliability \cite{DBLP:journals/corr/abs-1811-02553}. For this paper, we will be using a graph-learning PPO agent, trained with a policy network consisting of a Graph Neural Networks (GNN). GNNs have been successfully implemented in a wide variety of task allocation, scheduling, and path planning problems \cite{Paul_ICRA, 9750805, kool2018attention} in the past few years. One of the main advantages of GNNs is their ability to use the structural information (local and global) of a problem formulated as graph-structured data, and are represented as graph embedding, node embeddings, or edge embeddings. In this work we implement a Graph Convolutional Network (GCN) \cite{zhang2019graph} for graph embeddings and a custom multi-layer perceptron (MLP) for transforming a final feature vector into a set of log probabilities. The RL parameters used for training the networks are mentioned in the table \ref{tab:A2C_params_Airsim}, and the network architecture used for the GRL agent is shown in figure \ref{fig:grlagentnetwork}. The GRL agent MLP consists of 2 layers of 128 and 64 neurons shown in \ref{fig:grlagentnetwork}. Additionally, The agent utilizes masking which will depend on the state of the selected E-VTOL and the availability of each port. This takes away a layer of complexity and allows the agent to focus on other environmental factors, such as avoiding collisions and reducing uncertainty. As shown in figure \ref{fig:grlagentnetwork}, the GCN agent has a feature abstraction, policy and value network to work with PPO. We make use of biases for the linear layers and use randomized ReLU (RReLU) with a slope ranging from  $0.1$ to $0.3$, as this was the quickest and most effective option for mitigating vanishing gradients. Initially, we went with LeakyReLU, however, the time spent tuning the activation layer for each network was very time-consuming. The Adam optimizer with a learning rate of 1e-5 is used for back-propagation. The main difference lies in the feature abstraction network where we use two GCNs, which take the E-VTOL and vertiport feature matrix along with their respective edge connectivity matrix. The policy network will use a four-layer MLP with a log-softmax transformation to obtain log probabilities for the 11 actions. The agent utilizes masking which will depend on the state of the selected E-VTOL and the availability of each port. This takes away a layer of complexity and allows the agent to focus on other environmental factors, such as avoiding collisions and reducing uncertainty.

% \begin{figure}
%      \centering
%      \begin{subfigure}[b]{0.49\linewidth}
%          \centering
%          \includegraphics[width=\linewidth]{Figs/RL_Network_Flow.png}
%          \caption{}
%          \label{fig:rlagentnetwork}
%      \end{subfigure}
%      \hfill
%      \begin{subfigure}[b]{0.49\linewidth}
%          \centering
%          \includegraphics[width=\linewidth]{Figs/GRL_Network_Flow.png}
%          \caption{}
%          \label{fig:grlagentnetwork}
%      \end{subfigure}
%      \hfill
%      \caption{A Visualization of the (a) Baseline RL Policy Architecture and the (b) GRL Policy Architecture}
%      \label{fig:networkflow}
% \end{figure}

% \subsubsection{Baseline Agent}

% The vertiport states for baseline RL is modified to 3 different states, each indicating the availability of normal port, battery port, and hovering spot. Together with the VTOL states the total state space constitutes of 8 multi-discrete state spaces. The action space will have discrete values from 0 to 4 depending on the status of the vehicle. For instance, if the vehicle is in-air then the actions related to landing are considered, whereas if the status of vehicle is in the vertiport the actions related to the Takeoff are considered. 

% \subsubsection{Graph Learning Agent}
% For the graph learning approach, the graph properties of vertiports and VTOLs mentioned in table \ref{tab:MDP} are passed through the GCN feature abstractor and then to the MLP. The policy network used here also follows the architecture mentioned in section \ref{sec:RL}

% \subsubsection{Uncertainties}
% Real life uncertainties such as localization error and environmental stochasticity are added to the simulation. The  errors can be caused due to various reasons such as wind and communication loss. The effect is added in such a way that whenever an action is taken, the EVTOLs error in localization will be a normal distribution with $
% \mu = 1$ and $\sigma=1$, further 5\% of the time, the action is misinterpreted for a different action. 

% \usepackage{multirow}

% \usepackage{multirow}

% \usepackage{multirow}

% \usepackage{multirow}

\section{Simulation Environment}
\label{modelling}

\begin{figure*}[!ht]
\centering
\includegraphics[width=0.95\linewidth]{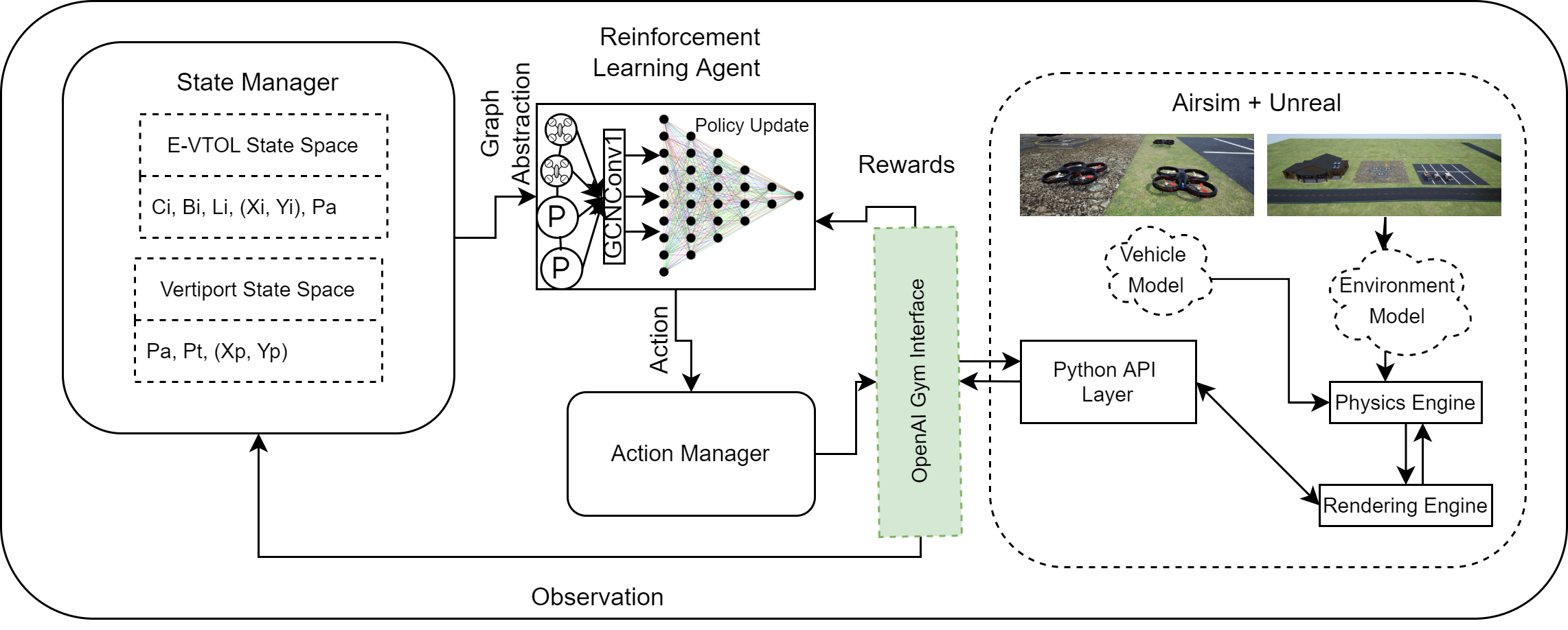}
\caption{Simulation Environment Overview Showing the Flow of Information through the State Manager, Action Manager, Learning Policy Network, and the Simulation Engine (Airsim + Unreal)}
\label{figs:flow}\vspace{-0.2cm}
\end{figure*}

A custom simulation for eVTOLs is developed on top of Microsoft AirSim. AirSim \cite{shah2017airsim} is an open-source robotics simulation platform. AirSim helps us to solve the need for large data sets for training and allows debugging in the simulator. AirSim leverages current game engines(Unreal Engine)\cite{sanders2016introduction} rendering, physics, and perception computation to create accurate, real-world simulations. Together, this realism, based on efficiently generated ground-truth data, enables the study and execution of complex, time-consuming, and risky missions in the real world. AirSim enables us to simulate the physics of eVTOLs, while the properties of vertiport and eVTOLs are programmatically implemented in \textit{Python}. The overall framework used for learning is shown in Figure \ref{figs:flow}. The OpenAI Gym-based Reinforcement Learning interface is developed for communication with the AirSim and the learning agent. We will now describe each component in more detail.

\subsection{OpenAI Gym Interface}
Gym is an open-source \textit{Python} library and it provides a standard API to communicate between learning algorithms and environments\cite{brockman2016openai}. Since its release, Gym API become the field standard for training and developing RL problems. In our case, the Gym interface plays a crucial role in communicating between all the different components of the learning framework. For instance, the Gym interface receives a decoded action from the action manager and it sends this information to the \textit{Python} API layer of AirSim, which in turn simulates the physics and dynamics of the vehicle.

\subsection{AirSim+Unreal }
\begin{figure}[!ht]
\centering
\includegraphics[width=0.9\linewidth]{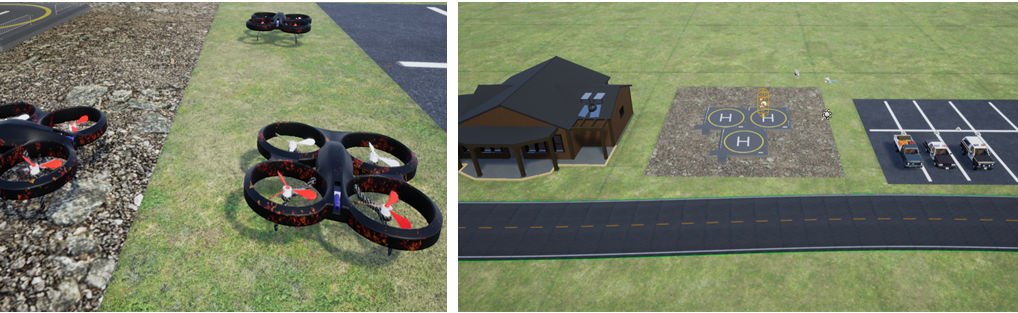}
\caption{The Simulation Environment Developed using Unreal Engine.}
\label{figs:env}\vspace{-0.2cm}
\end{figure}

A custom environment is built on Unreal Engine \cite{sanders2016introduction} with 3 vertiports and the AirSim plugin is incorporated with 5 VTOLs. The AirSim plugin manages the physics of the VTOLs and sends data such as location, and collision information to the OpenAI Gym interface. AirSim provides an option to run the simulation much faster than the real-world clock, this helps to speed up the training process. The AirSim API layer receives instructions from the Gym interface such as the go-to location/takeoff/land, the AirSim manages the physics and path planning while the unreal engine renders them on screen. The graphical interface is shown in figure \ref{figs:env}

\subsection{State Manager}
The state manager inherits all the properties of ports and VTOLs and is responsible for extracting the required state information of particular VTOLs requested by the Gym interface. The properties of VTOLs include 1. Battery level, 2. Schedule, 3. current position, 4. Status(on-time, delayed). We defined 2 \textit{Python} classes one for UAM and another for ports. Each VTOL will derive the properties of UAM class and the ports class encapsulates the properties of the vertiports. The ports class manages the vertiport and is responsible for sending out the status of individual ports to all the vehicles and updating the status when a vehicle lands or leaves the port. Two kinds of ports are considered here 1. Battery ports, only these ports are capable of charging the vehicles, and 2. Normal ports. The hover spots are also managed by ports class, the vehicles take one of these positions when entering the port zone and hover here till the agent decides an action. 

\subsection{Action Manager}
The Action manager inherits all the properties of ports and VTOLs and is responsible for decoding the actions sent by the RL agent. Once decoded the action is sent to the vehicle and the status of vehicle and port are updated accordingly. The Gym interface communicates between the RL agent and the action manager. 
\begin{table}[htbp]
\small
\centering
% To place a caption above a table
\caption{Reinforcement Learning Parameters}
\begin{tabular}[t]{cc}
    \toprule
% Table content
    % \hline
         Algorithm & PPO  \\
         
         \midrule
        %  \hline
        %  Policy & MLP Policy  \\
        %  Policy Network & MLP/GCN   \\
        %  \hline
         Maximum Timesteps & 300,000  \\
        %  \hline
         Learning Rate & 1e-5 \\
        %  \hline
         Discount Factor & 1 \\
        %  \hline
         Number of Steps & 20,000 \\
        %  \hline
        Batch Size & 10,000 \\
        %   \hline
         Entropy Coefficient & 0.001 \\
        %  \hline
        Reward Weights & $\{0.3, 0.3, 0.35, 0.1, 0.35\}$ \\
        %   \hline
    \bottomrule
\end{tabular}
% Or to place a caption below a table
% \caption{Test}
\label{tab:A2C_params_Airsim}
\end{table}%

\section{Results and Discussion}
\label{sec:Results}

This section is split into two sub-sections: a learning section and a case study section. The learning section will go over the training: what went well, and what we still need to work on. The case study section will focus more on evaluating the GRL agent with and without uncertainty in the environment. Here uncertainty means three things:

\begin{enumerate}
    \item Wind effects are added to the environment as an adversary vector, which will lower the linear velocity of the eVTOL or negate it completely with a 5\% occurrence.
    \item Battery ports have a 5\% chance of not working on a given step.
    \item eVTOLs will have a 5\% chance of not taking off.
\end{enumerate}

These three changes to the environment will simulate inclement weather, faulty equipment and mechanical failures of eVTOLs respectively. Ideally, the agent will learn how to account for these changes by: sending eVTOLs off earlier to their destinations to account for wind effect slowdowns, keeping eVTOLs charging for an extra step if their charging equipment is faulty, and takeoff immediately once mechanical failures are addressed to save time. During this, the agent is still expected to uphold the highest level of safety for passengers, so it will be heavily penalized for collisions. The agent will be tested for 50 episodes (72000 steps) in each case study, with each episode representing 24 hours or 1 day of vertiport operation. Then, it will be compared against a random agent and a first come first serve (FCFS) agent. The random agent is quite simple in nature, and takes a random action at each step, without taking the state space into consideration. On the other hand, the FCFS agent will use queues to decide which eVTOL should take off and which should land and charge. While an eVTOL is waiting in a queue, it will wait in a normal port, or a hovering spot if the normal ports are taken. The FCFS agent will also charge each eVTOL until it is at or above 60\%, after which it will take off and enter the landing queue.

\subsection{Learning Curve}\label{sec:learning_curve}
% The GRL agent was trained for 1 million steps, which comes out to 694 episodes. Figure \ref{figs:rewardloss} shows the reward and loss for a model trained with and without noise in the environment\footnote{These results are not indicative of the models' true performance}. Due to complications during training, we were not able to obtain the ideal results to showcase, as the parameter space is large and we had limited time to produce these results with the equipment at our disposal. It is clear both models were able to learn, until there was a sudden drop in performance. This is due to vanishing gradients, which affect deep learning problems of all scales. This can happen when: {\bf i.)} there are too many layers in a network, causing already small weights to become infinitesimally small and disrupt the gradient descent during back-propagation {\bf ii.)} The activation function may aggressively approximate smaller weights to 0, such as the sigmoid or ReLU nonlinearities, and as such negative weights will get suddenly ignored, causing instability during updates. {\bf iii.)} The learning rate of the reinforcement learning algorithm may be too small, which can make the policy gradient 0 through smaller weight updates. These are all problems we tried to address through hyper-parameter tuning and modifying the value and policy networks. Once this problem is solved, we believe the models will both be able to learn their respective environment.

\begin{figure}[!ht]
\centering
\includegraphics[width=0.9\linewidth]{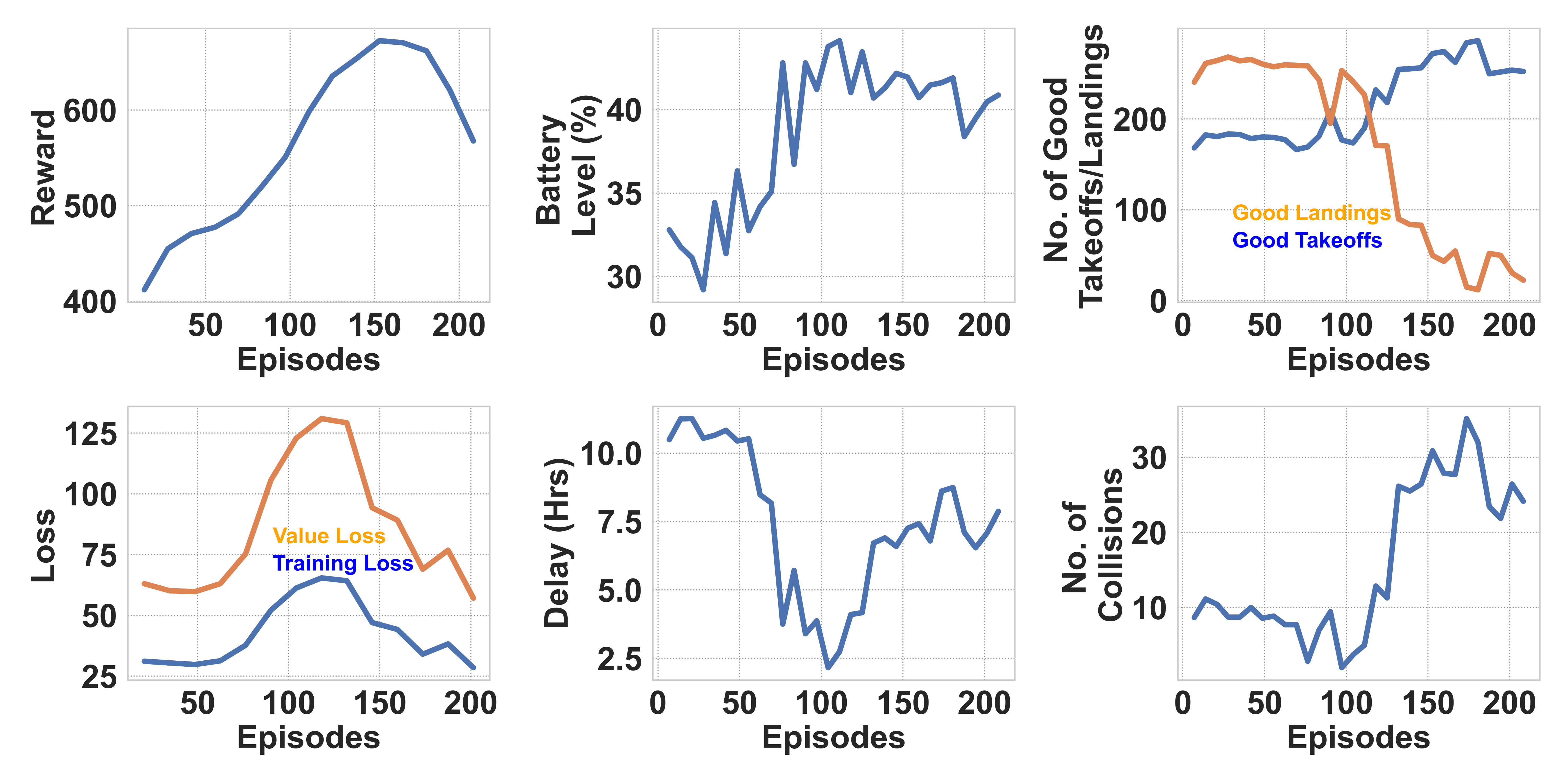}
\caption{Reward and loss plot for the GRL agent along with environment metrics for interpretability}
\label{figs:rewardloss}\vspace{-0.2cm}
\end{figure}

The GRL agent was trained for 300k steps, or 200 episodes and the training plots are included in figure \ref{figs:rewardloss}. From the reward and loss plots, we can tell that the agent was able to generalize the environment, for the reward was steadily increasing and the loss was steadily decreasing. There was a sudden increase in the loss of around 60 episodes, which indicates the agent wasn't making the right approximations with the state space information it received, although the loss eventually went back down and continued to decrease. Additionally, we include other plots in figure \ref{figs:rewardloss} which show: the average battery levels of each drone per episode; the number of good takeoffs and good landings by the agent per episode; the average delay of a single drone in hours per episode, and the number of collisions per episode the agent experiences. This information was recorded during training and helps us interpret the agent's performance. We can see that the agent starts off with a lower battery level, and eventually learns to keep it over 30\% at all times, as anything lower will lead to reward penalties. The delay also starts off very high, but as the agent gets used to taking off and landing it starts to diminish as well. Unfortunately, the agent never quite learns how to avoid all collisions, as it avoids some but incurs more as time goes on. This is due to the weights for safety not being high enough. If the reward weight for safety was higher, the agent would be forced to minimize all collisions in order to maintain a high reward, instead of trying to offset the collision penalty with more good takeoffs and landings. 

\subsection{Case Study(s)}\label{sec:case_studies}

\begin{figure*}[!ht]
\centering
\includegraphics[width=0.9\linewidth]{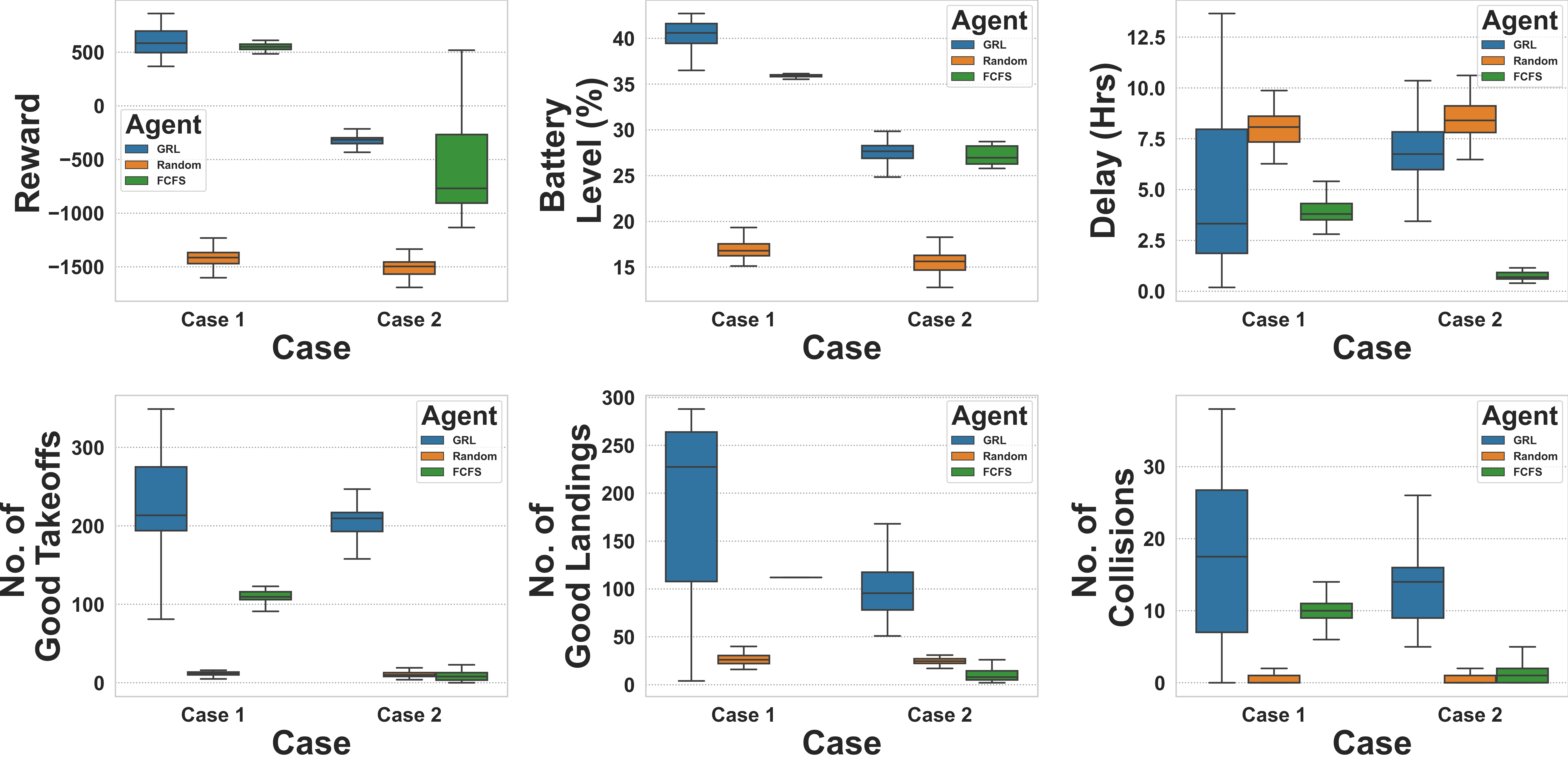}
\caption{Case study results. In case 1 the three agents are tested without noise in the environment, and in case 2 they are tested with noise.}
\label{figs:casestudy}\vspace{-0.2cm}
\end{figure*}

For the case studies, we chose to use the ideal trained GRL model at 160k steps, or 111 episodes. At that point in training, the collisions were the lowest, the good takeoffs and landings were balanced, the delay was close to a local minimum, and the average battery levels were close to a local max as shown in figure \ref{figs:rewardloss}. The case study results can be found in figure \ref{figs:casestudy}. \\
{\bf Case 1:} the GRL agent does well across every metric except for collisions. Noticeably, the GRL agent has more deviation in every metric, which is likely due to the agent adapting to various situations per episode. If the agent had more time to train and fine-tune it's policy, these deviations would be smaller and more controlled. The FCFS agent was the second-best performer, and was more consistent along each metric. The random agent performed the worst out of the three, which is expected. \\
{\bf Case 2:} The performance across the GRL and FCFS agents decreased substantially in some metrics and less so in others. The GRL agent reacts as expected to the noisy additions in the environment: the reward drops, indicating some difficulty maximizing performance across all metrics; the battery levels drop by 13\%, a direct cause of the charging ports not malfunctioning; the delay also increases while the number of good takeoffs decreases, the effect of harsh wind and eVTOL mechanical failures during takeoff; good landings decrease as well due to eVTOLs no longer being on time to land; lastly, the number of collisions decrease by a small percentage, due to a direct correlation with eVTOLs taking off less. The FCFS agent also sees a decrease in performance for the same reasons, while the random agent is far behind the other two baselines in every metric except collisions. Interestingly, the FCFS agent has less delay overall when noise is present, and this is likely due to it's consistency when completing tasks (taking off an eVTOL and then landing it in two queues) which keeps giving eVTOLs new schedules and resetting their delay. If an eVTOL is in the queue but not at the front, the FCFS agent will skip over it, which saves time and minimizes delay. On the other hand, the GRL agent will wait for an eVTOL to be in it's airspace to take an action for it, which can lead to deceptively increased delay times. \\
These results show the GRL agent can actually learn the parameter space of our simulated environment, and furthermore adapt to added noise in the simulation with varying levels of success. Given more time to train and tune hyper-parameters, and a better combination of reward weights, we believe this agent can do well in every metric, with and without noise present in the environment. 

\subsection{Action Analysis}
\begin{figure*}[!ht]
\centering
\includegraphics[width=0.65\linewidth]{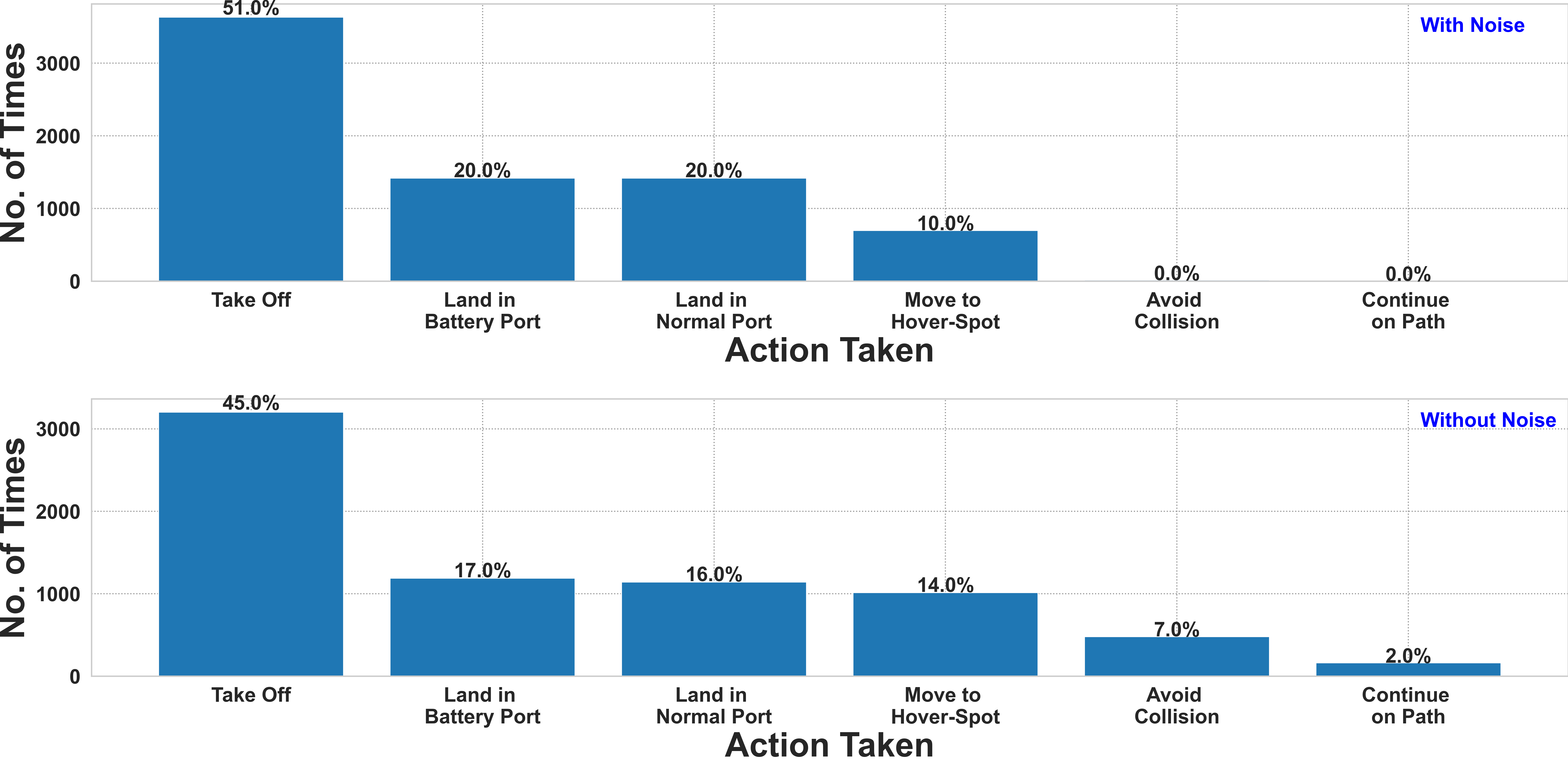}
\caption{Action distribution for the GRL agent in an environment with and without noise accumulated over 5 episodes}
\label{figs:actoindis}
\end{figure*}

Additionally, we recorded the GRL agent in the simulation environment with and without noise for 5 episodes each, and charted the action distribution in figure \ref{figs:actoindis}. This chart shows the count of each action the agent chose during those 5 episodes, along with the percentage of time those actions were chosen. Right away we notice the agent is not using all of the actions it's been allotted, which either means it doesn't deem them necessary for maximizing the reward or that the agent hasn't had enough time to figure out how to use them properly. Another reason why the agent might not need to use all of the actions is due to the small group of eVTOLs it's controlling. it will be interesting to see how the agent scales up, and how it will use more than one port, and one hovering spot to keep up it's current rhythm. The second important thing to note has to do with the agent distribution with noise. The actions chosen are the same, however the frequency changes, especially for avoiding collision and continue actions. This shows the agent tried to make changes in real-time to adapt to the noise; it could be that it tried to avoid a collision due to inclement wind, or something similar. This also speaks to the effectiveness of the simulator we chose. The wind vectors we created were realistic enough to present a real challenge to the agent, and in doing so we allowed it to further generalize the environment and growing parameter space as a direct product of the added noise. 

\section{Conclusion}\label{sec:Conclusion}
In this paper, we proposed a graph-based reinforcement learning (RL) for Urban Air Mobility (UAM) vertiport operations management problem in the presence of various environmental uncertainties (wind gust effects), and observational uncertainties (malfunctioning battery ports and eVTOL take-off delay). The novelty of our approach lies in the introduction of Graph Neural Networks (GNN), which serve as a feature abstraction for the state space, for the RL framework. The reward function for the MDP was tailored as a weighted sum of terms that quantify good take-offs, good landings, battery state, delay, and safety. Initially, the UAM vertiport operations management problem is formulated as a Markov Decision Process (MDP), and a policy-gradient RL method  (Proximal Policy Optimization) was used to solve the MDP. The GNN-based policy network consists of two Graph Convolutional Networks (GCN), which are used to encode the vertiport state information and the eVTOL state information (both represented as a graph), respectively. The RL environment was modeled as an OpenAI gym-based environment using AirSim and Unreal engine. The trained policy (best performing) was tested on two sets of test scenarios. In the first set, the test scenarios were drawn from the same distribution as for training, while in the second set the test scenarios had uncertainty. For baseline comparison, a first come first serve (FCFS), and a random agent was used.  Both sets of scenarios consist of 50 unseen episodes. From both cases, the GRL agent has a clear advantage in comparing w.r.t the total rewards. On a closer look at the individual metrics that constitute the reward, it can be seen that the GRL agent performs poorly only for the number of collisions. This can be mitigated by prioritizing the number of collisions by increasing the weight corresponding to it while training. The two case studies demonstrate the ability of the GRL agent to generalize across unseen scenarios and with uncertainty. Further analyses of the actions taken show how the uncertainties altered the decision-making to take less take-off and more collision avoidance and hence to be more conservative.

% \newpage
\bibliographystyle{IEEEtran}

\bibliography{sample}

\begin{thebibliography}{10}
\providecommand{\url}[1]{#1}
\csname url@samestyle\endcsname
\providecommand{\newblock}{\relax}
\providecommand{\bibinfo}[2]{#2}
\providecommand{\BIBentrySTDinterwordspacing}{\spaceskip=0pt\relax}
\providecommand{\BIBentryALTinterwordstretchfactor}{4}
\providecommand{\BIBentryALTinterwordspacing}{\spaceskip=\fontdimen2\font plus
\BIBentryALTinterwordstretchfactor\fontdimen3\font minus
  \fontdimen4\font\relax}
\providecommand{\BIBforeignlanguage}[2]{{%
\expandafter\ifx\csname l@#1\endcsname\relax
\typeout{** WARNING: IEEEtran.bst: No hyphenation pattern has been}%
\typeout{** loaded for the language `#1'. Using the pattern for}%
\typeout{** the default language instead.}%
\else
\language=\csname l@#1\endcsname
\fi
#2}}
\providecommand{\BIBdecl}{\relax}
\BIBdecl

\bibitem{united_nations_2018}
\BIBentryALTinterwordspacing
``68\% of the world population projected to live in urban areas by 2050, says
  un | un desa department of economic and social affairs,'' May 2018. [Online].
  Available:
  \url{https://www.un.org/development/desa/en/news/population/2018-revision-of-world-urbanization-prospects.html}
\BIBentrySTDinterwordspacing

\bibitem{rothfeld2020urban}
R.~Rothfeld, A.~Straubinger, M.~Fu, C.~Al~Haddad, and C.~Antoniou, ``Urban air
  mobility,'' in \emph{Demand for Emerging Transportation Systems}.\hskip 1em
  plus 0.5em minus 0.4em\relax Elsevier, 2020, pp. 267--284.

\bibitem{BAURANOV2021100726}
\BIBentryALTinterwordspacing
A.~Bauranov and J.~Rakas, ``Designing airspace for urban air mobility: A review
  of concepts and approaches,'' \emph{Progress in Aerospace Sciences}, vol.
  125, p. 100726, 2021. [Online]. Available:
  \url{https://www.sciencedirect.com/science/article/pii/S0376042121000312}
\BIBentrySTDinterwordspacing

\bibitem{horne2019next}
T.~A. HORNE, ``Next up: evtols: Uber leads the way into a new, jetsons-like
  future,'' \emph{AOPA Pilot}, 2019.

\bibitem{vascik2017constraint}
P.~D. Vascik and R.~J. Hansman, ``Constraint identification in on-demand
  mobility for aviation through an exploratory case study of los angeles,'' in
  \emph{17th AIAA Aviation Technology, Integration, and Operations Conference},
  2017, p. 3083.

\bibitem{daskilewicz2018progress}
M.~Daskilewicz, B.~German, M.~Warren, L.~A. Garrow, S.-S. Boddupalli, and T.~H.
  Douthat, ``Progress in vertiport placement and estimating aircraft range
  requirements for evtol daily commuting,'' in \emph{2018 Aviation Technology,
  Integration, and Operations Conference}, 2018, p. 2884.

\bibitem{guerreiro2020capacity}
N.~M. Guerreiro, G.~E. Hagen, J.~M. Maddalon, and R.~W. Butler, ``Capacity and
  throughput of urban air mobility vertiports with a first-come, first-served
  vertiport scheduling algorithm,'' in \emph{AIAA Aviation 2020 Forum}, 2020,
  p. 2903.

\bibitem{preis2022vertiport}
L.~Preis and M.~Hornung, ``Vertiport operations modeling, agent-based
  simulation and parameter value specification,'' \emph{Electronics}, vol.~11,
  no.~7, p. 1071, 2022.

\bibitem{bacchini2019electric}
A.~Bacchini and E.~Cestino, ``Electric vtol configurations comparison,''
  \emph{Aerospace}, vol.~6, no.~3, p.~26, 2019.

\bibitem{li2017deep}
Y.~Li, ``Deep reinforcement learning: An overview,'' \emph{arXiv preprint
  arXiv:1701.07274}, 2017.

\bibitem{https://doi.org/10.48550/arxiv.1911.10635}
\BIBentryALTinterwordspacing
K.~Zhang, Z.~Yang, and T.~Başar, ``Multi-agent reinforcement learning: A
  selective overview of theories and algorithms,'' 2019. [Online]. Available:
  \url{https://arxiv.org/abs/1911.10635}
\BIBentrySTDinterwordspacing

\bibitem{Kool2019}
W.~Kool, H.~{Van Hoof}, and M.~Welling, ``{Attention, learn to solve routing
  problems!}'' in \emph{7th International Conference on Learning
  Representations, ICLR 2019}, 2019.

\bibitem{barrett2019exploratory}
T.~D. Barrett, W.~R. Clements, J.~N. Foerster, and A.~I. Lvovsky, ``Exploratory
  combinatorial optimization with reinforcement learning,'' \emph{arXiv
  preprint arXiv:1909.04063}, 2019.

\bibitem{doi:10.2514/6.2022-3911}
\BIBentryALTinterwordspacing
S.~Paul and S.~Chowdhury, \emph{A Graph-based Reinforcement Learning Framework
  for Urban Air Mobility Fleet Scheduling}. [Online]. Available:
  \url{https://arc.aiaa.org/doi/abs/10.2514/6.2022-3911}
\BIBentrySTDinterwordspacing

\bibitem{khalil2017learning}
E.~Khalil, H.~Dai, Y.~Zhang, B.~Dilkina, and L.~Song, ``Learning combinatorial
  optimization algorithms over graphs,'' in \emph{Advances in Neural
  Information Processing Systems}, 2017, pp. 6348--6358.

\bibitem{Kaempfer2018LearningTM}
Y.~Kaempfer and L.~Wolf, ``Learning the multiple traveling salesmen problem
  with permutation invariant pooling networks,'' \emph{ArXiv}, vol.
  abs/1803.09621, 2018.

\bibitem{9750805}
R.~A. Jacob, S.~Paul, W.~Li, S.~Chowdhury, Y.~R. Gel, and J.~Zhang,
  ``Reconfiguring unbalanced distribution networks using reinforcement learning
  over graphs,'' in \emph{2022 IEEE Texas Power and Energy Conference (TPEC)},
  2022, pp. 1--6.

\bibitem{li2018combinatorial}
Z.~Li, Q.~Chen, and V.~Koltun, ``Combinatorial optimization with graph
  convolutional networks and guided tree search,'' in \emph{Advances in Neural
  Information Processing Systems}, 2018, pp. 539--548.

\bibitem{nowak2017note}
A.~Nowak, S.~Villar, A.~S. Bandeira, and J.~Bruna, ``A note on learning
  algorithms for quadratic assignment with graph neural networks,''
  \emph{stat}, vol. 1050, p.~22, 2017.

\bibitem{Tolstaya2020MultiRobotCA}
E.~V. Tolstaya, J.~Paulos, V.~R. Kumar, and A.~Ribeiro, ``Multi-robot coverage
  and exploration using spatial graph neural networks,'' \emph{ArXiv}, vol.
  abs/2011.01119, 2020.

\bibitem{Sykora2020}
Q.~Sykora, M.~Ren, and R.~Urtasun, ``{Multi-agent routing value iteration
  network},'' in \emph{37th International Conference on Machine Learning, ICML
  2020}, 2020.

\bibitem{paul_cvrp}
\BIBentryALTinterwordspacing
S.~Paul and S.~Chowdhury, ``A scalable graph learning approach to capacitated
  vehicle routing problem using capsule networks and attention mechanism,'' in
  \emph{International Design Engineering Technical Conferences and Computers
  and Information in Engineering Conference}, vol. Volume 3B: 48th Design
  Automation Conference (DAC), 08 2022, v03BT03A045. [Online]. Available:
  \url{https://doi.org/10.1115/DETC2022-90123}
\BIBentrySTDinterwordspacing

\bibitem{Dai2017}
H.~Dai, E.~B. Khalil, Y.~Zhang, B.~Dilkina, and L.~Song, ``{Learning
  combinatorial optimization algorithms over graphs},'' in \emph{Advances in
  Neural Information Processing Systems}, 2017.

\bibitem{Paul_ICRA}
S.~Paul, P.~Ghassemi, and S.~Chowdhury, ``Learning scalable policies over
  graphs for multi-robot task allocation using capsule attention networks,'' in
  \emph{2022 International Conference on Robotics and Automation (ICRA)}, 2022,
  pp. 8815--8822.

\bibitem{hernandez2018multiagent}
P.~Hernandez-Leal, B.~Kartal, and M.~E. Taylor, ``Is multiagent deep
  reinforcement learning the answer or the question? a brief survey,''
  \emph{learning}, vol.~21, p.~22, 2018.

\bibitem{behjat2021learning}
A.~Behjat, H.~Manjunatha, P.~K. Kumar, A.~Jani, L.~Collins, P.~Ghassemi,
  J.~Distefano, D.~Doermann, K.~Dantu, E.~Esfahani \emph{et~al.}, ``Learning
  robot swarm tactics over complex adversarial environments,'' in \emph{2021
  International Symposium on Multi-Robot and Multi-Agent Systems (MRS)}.\hskip
  1em plus 0.5em minus 0.4em\relax IEEE, 2021, pp. 83--91.

\bibitem{lyu2021contrasting}
X.~Lyu, Y.~Xiao, B.~Daley, and C.~Amato, ``Contrasting centralized and
  decentralized critics in multi-agent reinforcement learning,'' \emph{arXiv
  preprint arXiv:2102.04402}, 2021.

\bibitem{schulman2017proximal}
J.~Schulman, F.~Wolski, P.~Dhariwal, A.~Radford, and O.~Klimov, ``Proximal
  policy optimization algorithms,'' \emph{arXiv preprint arXiv:1707.06347},
  2017.

\bibitem{https://doi.org/10.48550/arxiv.1502.05477}
\BIBentryALTinterwordspacing
J.~Schulman, S.~Levine, P.~Moritz, M.~I. Jordan, and P.~Abbeel, ``Trust region
  policy optimization,'' 2015. [Online]. Available:
  \url{https://arxiv.org/abs/1502.05477}
\BIBentrySTDinterwordspacing

\bibitem{openai}
G.~Brockman, V.~Cheung, L.~Pettersson, J.~Schneider, J.~Schulman, J.~Tang, and
  W.~Zaremba, ``Openai gym,'' 2016.

\bibitem{stable-baselines}
A.~Hill, A.~Raffin, M.~Ernestus, A.~Gleave, A.~Kanervisto, R.~Traore,
  P.~Dhariwal, C.~Hesse, O.~Klimov, A.~Nichol, M.~Plappert, A.~Radford,
  J.~Schulman, S.~Sidor, and Y.~Wu, ``Stable baselines,''
  \url{https://github.com/hill-a/stable-baselines}, 2018.

\bibitem{DBLP:journals/corr/abs-1811-02553}
\BIBentryALTinterwordspacing
A.~Ilyas, L.~Engstrom, S.~Santurkar, D.~Tsipras, F.~Janoos, L.~Rudolph, and
  A.~Madry, ``Are deep policy gradient algorithms truly policy gradient
  algorithms?'' \emph{CoRR}, vol. abs/1811.02553, 2018. [Online]. Available:
  \url{http://arxiv.org/abs/1811.02553}
\BIBentrySTDinterwordspacing

\bibitem{kool2018attention}
\BIBentryALTinterwordspacing
W.~Kool, H.~van Hoof, and M.~Welling, ``Attention, learn to solve routing
  problems!'' in \emph{International Conference on Learning Representations},
  2019. [Online]. Available: \url{https://openreview.net/forum?id=ByxBFsRqYm}
\BIBentrySTDinterwordspacing

\bibitem{zhang2019graph}
S.~Zhang, H.~Tong, J.~Xu, and R.~Maciejewski, ``Graph convolutional networks: a
  comprehensive review,'' \emph{Computational Social Networks}, vol.~6, no.~1,
  pp. 1--23, 2019.

\bibitem{shah2017airsim}
S.~Shah, D.~Dey, C.~Lovett, and A.~Kapoor, ``Airsim: High-fidelity visual and
  physical simulation for autonomous vehicles,'' in \emph{Field and service
  robotics}.\hskip 1em plus 0.5em minus 0.4em\relax Springer, 2018, pp.
  621--635.

\bibitem{sanders2016introduction}
A.~Sanders, \emph{An introduction to Unreal engine 4}.\hskip 1em plus 0.5em
  minus 0.4em\relax AK Peters/CRC Press, 2016.

\bibitem{brockman2016openai}
G.~Brockman, V.~Cheung, L.~Pettersson, J.~Schneider, J.~Schulman, J.~Tang, and
  W.~Zaremba, ``Openai gym,'' \emph{arXiv preprint arXiv:1606.01540}, 2016.

\end{thebibliography}

\end{document}